\documentclass[showpacs,preprintnumbers,superscriptaddress,amsmath,amssymb]{revtex4-2}

\usepackage[colorlinks=true,bookmarks=false,citecolor=red,urlcolor=blue,linkcolor=blue]{hyperref}
\usepackage[top=2.5cm, bottom=2.5cm,left=3cm, right=3cm]{geometry}
\usepackage{amsmath,amssymb}
\usepackage{graphicx}
\usepackage{verbatim}
\usepackage{enumitem}
\usepackage{color}
\usepackage[english]{babel}
\usepackage{bm}
\usepackage{natbib}
\usepackage[symbol]{footmisc}
\usepackage{textcomp}
\usepackage{epstopdf}
\usepackage{setspace}

\def\opex{Opt. Express}
\def\ol{Opt. Lett.}

\def\pra{Phys. Rev. A}
\def\prb{Phys. Rev. B}

\def\prl{Phys. Rev. Lett.}
\def\nc{Nat. Commun.}

\def\siamjam{SIAM J. Appl. Math.}

\def\bp{{\bf\Phi}}
\def\cbp{\overline{\bp}}
\def\bs{{\bf\Psi}}

\def\bE{{\bf E}}

\def\1e{\hat{x}}
\def\2e{\hat{y}}
\def\3e{\hat{z}}
\def\L{\mathcal{L}}
\def\B{\mathcal{B}}
\def\P{\mathcal{P}}
\def\T{\mathcal{T}}
\def\S{\mathcal{S}}

\def\rmd{{\rm d}}
\def\br{{\bf r}}
\def\dr{\rmd\br}
\def\mum{\mu\mbox{m}}

\begin{document}

\title{Non-generic bound states in the continuum in
	waveguides with lateral leakage channels}

\author{Nan Zhang}
\thanks{Corresponding Author}
\email{nzhang234-c@my.cityu.edu.hk} 
\affiliation{Department of Mathematics, City University of Hong Kong, Hong Kong}

\author{Ya Yan Lu}
\affiliation{Department of Mathematics, City University of Hong Kong, Hong Kong}
\date{\today}

\begin{abstract}
	For optical waveguides with a layered background which itself is a slab waveguide,
	a guided mode is a bound state in the continuum (BIC), if it coexists with {slab modes}
	propagating outwards in the lateral direction;   i.e., there are lateral leakage channels.
	It is known that generic BICs in optical waveguides with lateral leakage channels are robust
	in the sense that they still exist if the waveguide is perturbed arbitrarily.
	However, the theory is not applicable to non-generic BICs
	which can be defined precisely. Near a BIC, the waveguide supports resonant and leaky modes
	with a complex frequency and  a complex propagation constant, respectively.
	In this paper, we develop a perturbation theory to show that the resonant and leaky modes 
	near a non-generic BIC have an ultra-high $Q$ factor and ultra-low leakage loss, respectively.
	We also show that a {\em merging}-BIC obtained by tuning structural parameters
	is always a non-generic BIC.
	Existing studies on {\em merging}-BICs are concerned with specific examples and specific parameters.
	We analyze an arbitrary structural perturbation (to a waveguide supporting a non-generic BIC) given by $\delta F(\br)$,
	where $F(\br)$ is the perturbation profile and $\delta$ is the amplitude, and show that
	the perturbed waveguide has two BICs for $\delta>0$ (or $\delta<0$) and no BIC for $\delta<0$ (or $\delta>0$).
	This implies that a non-generic BIC is a {\em merging}-BIC
	(for any perturbation profile $F$) when $\delta$ is regarded as a parameter. 
	Our study indicates that non-generic BICs have interesting
	special properties that are useful in applications.
\end{abstract}

\maketitle

\section{Introduction}

Some optical waveguides, such as the strip or ridge waveguides, consist
of a core in a layered background which itself is a planar waveguide (usually, a slab
waveguide)~\cite{snyder83,Peng78IEICE,Peng81MTT1,Oliner81MTT2}.
Such a waveguide may have only leaky modes for which power is lost
laterally by coupling with outgoing propagating modes of the background planar
waveguide~\cite{Peng78IEICE,Ogusu79AO,Ogusu80AO,Peng81MTT1,Oliner81MTT2,Ogusu83JOSA}. It
has been observed long time ago that by tuning the structural parameters, the
leakage loss of a leaky mode in such a waveguide can be sharply 
reduced\cite{Peng78IEICE,Ogusu79AO,Ogusu80AO,Peng81MTT1,Oliner81MTT2,Ogusu83JOSA,webster07PTL,koshiba08OL}.
In fact, the leakage loss can be completely inhibited, and in that
case, the leaky mode becomes a bound state in the continuum 
(BIC)~\cite{zou15LPR,bezus18PR,yu19AOM,yu19Optica,Nguyen19LPR,Nguyen19JSTQE,yu20NC}.
More precisely, a BIC in such a waveguide with lateral leakage
channels (assuming there is no material loss) is a true guided mode
with a real angular frequency $\omega$, a real
propagation constant $\beta$, and a field confined around the core,
but $\beta$ is less than the largest propagation constant $\eta_{\rm max}$ of all propagating modes of the
background planar waveguide. Notice that the BIC coexists with a
propagating mode of the background planar waveguide having the
in-plane wavevector $(\pm \alpha_{\rm max}, \beta)$, where
$\alpha_{\rm max} = ( \eta^2_{\rm max} - \beta^2)^{1/2}>0$. A scattering
problem can be formulated with the above propagating mode serving as
incoming and outgoing waves. The existence of a BIC implies that the scattering
problem does not have a unique solution.

Photonic BICs exist  in many different
structures~\cite{Neumann1929PZ,Friedrich85PRA,Hsu16NRM,Koshelev19Nano,Sadreev21RPP,Azzam21AOM,Joseph21Nano}, 
and have found useful applications in lasing,
sensing~\cite{Romano18Material,Jacobsen22ACS},
switching~\cite{Han19AM}, nonlinear optics~\cite{Carletti18PRL,Yuan20JAM}, etc.  
For lossless structures with a single invariant or periodic direction, a BIC is
associated with a real frequency and a real propagation constant (or
Bloch wavenumber), and it is often regarded as a special
member in a continuous family of resonant or leaky modes. 
Both resonant and leaky modes are eigenmodes satisfying outgoing
radiation conditions. They are defined for a real $\beta$ and a real
$\omega$, and have a complex $\omega$ and a complex
$\beta$, respectively. The families of resonant and leaky
modes vary continuously with $\beta$ and $\omega$. Near
a typical BIC with frequency $\omega_*$ and propagation constant
$\beta_*$,  a resonant mode has a complex frequency with
$\mbox{Im}(\omega) \sim  |\beta-\beta_*|^2$ (quality factor $Q \sim 1/|\beta-\beta_*|^2$), and a leaky mode
has a complex propagation constant with $\mbox{Im}(\beta) \sim |\omega
- \omega_*|^2$.

For practical applications, it is important to understand how small
perturbations of the structure affect the BICs. 
In the perturbed structure, there is usually no BIC with the same
$\beta_*$ (if $\beta_* \ne 0$) or the same $\omega_*$. If the amplitude of the
perturbation is $\delta$, then the resonant mode with the same $\beta_*$
has $\mbox{Im}(\omega) \sim \delta^2$, and
the leaky mode   
with the same $\omega_*$  has
$\mbox{Im}(\beta) \sim \delta^2$~\cite{Kosh18PRL,Hu18JPB,amgad22PRA}.  
However, we can still ask whether there is a BIC in the perturbed
structure with a real pair $(\beta, \omega)$ near $(\beta_*, \omega_*)$.
A BIC is called robust with respect to a set of perturbations, if for
any sufficiently small  perturbation in that set, there is a BIC in the perturbed structure
with $(\beta, \omega)$ near $(\beta_*,
\omega_*)$. Symmetry protected BICs are clearly robust with respect to
symmetry-preserving perturbations, but BICs unprotected by symmetry
can also be robust~\cite{Yuan17OL,Yuan21PRACond}. 
In addition, if some tunable 
parameters are introduced in the perturbation, even a 
non-robust BIC can continue its existence in the perturbed
structure if the tunable parameters are properly chosen~\cite{Yuan20PRAPara,Yuan21PRAPara}. In fact, the minimum
number of tunable parameters needed is a unique integer for the BIC
and it is independent of the specific perturbations~\cite{2023pd3}. 

It is known that some BICs in optical waveguides with lateral leakage
channels are robust~\cite{bezus18PR, Bezus20Nano,Yuan21OE}. More
precisely, if the following three 
conditions are satisfied: (1) the waveguide has a lateral
mirror symmetry; (2) only one propagating mode of the background
planar waveguide has a propagation constant larger than that of the
BIC; (3) the BIC is {\em generic},  then the BIC is robust with respect to any
sufficiently small perturbation that preserves the lateral mirror
symmetry~\cite{Yuan21OE}. The first two conditions above ensure that there is
only one independent radiation channel. The third condition is given 
precisely in Section 2 and it  involves an integral related to the BIC and a corresponding
scattering solution.

In this paper, we study non-generic BICs in optical waveguides with
lateral leakage channels. It is assumed that conditions (1) and (2)
above are still satisfied, but the BIC is non-generic, namely,  the
integral mentioned above is zero. Since the BIC is surrounded by
resonant and leaky modes (for $\beta$ near $\beta_*$ and $\omega$ near
$\omega_*$,  respectively), we use a perturbation
method to show that typically, the nearby resonant and leaky modes have 
$\mbox{Im}(\omega) \sim (\beta-\beta_*)^4$ and 
$\mbox{Im}(\beta) \sim (\omega -\omega_*)^4$, respectively.
This implies that a resonant mode near a non-generic BIC has
an ultra-high quality factor ($Q$ factor), and a leaky mode near this
BIC has  ultra-low leakage loss. It should be mentioned that BICs
surrounded by resonant modes with an ultra-high $Q$ factor have been
found in many
studies~\cite{Yuan17PRA,Yuan18PRA,Yuan20PRAPert,Bulgakov17PRL,Zhen19Nature,Kang21PRL,Yuri21NC,kang22LSA,Bulgakov23PRB},
and they are referred to as  
{\em super}-BICs by some authors \cite{Yuri21NC,Bulgakov23PRB}. 
Moreover, a BIC surrounded by leaky modes with ultra-low 
leakage loss has been observed in an early work~\cite{koshiba08OL}.
Our theory reveals that a non-generic BIC is always a {\em
	super}-BIC.

The other purpose of this work is to find out whether 
BICs can persist under structural perturbations. The
existing theory on robustness covers only generic
BICs~\cite{Yuan21OE}.  Our study indicates that non-generic BICs
are indeed not robust, and the perturbed waveguide may or may not have
BICs. We consider a general perturbation to the 
dielectric function given by an arbitrary profile $F$ (that preserves  the lateral
mirror symmetry) multiplied by an amplitude $\delta$, and show that the perturbed
waveguide has no BIC for $\delta < 0$ (or $\delta>0$) and two BICs for
$\delta > 0$ (or $\delta < 0$). Since a pair of BICs split out of the
non-generic BIC, $\delta=0$ is the 
bifurcation point of a saddle node bifurcation~\cite{Smale12}. On the
other hand, as the positive (or negative
$\delta$) tends to 0, the two BICs merge to the non-generic BIC,
therefore, we can say that the non-generic BIC is a {\em merging}-BIC~\cite{Bulgakov17PRATopo,Bulgakov17PRL,Zhen19Nature,Kang21PRL,kang22LSA}.
In existing works on {\em merging}-BICs, one studies how two or more BICs
on a dispersion surface (or curve) of resonant modes merge
together as a structural parameter tends to a particular
value. The resulting BIC in the structure with that particular
parameter value is called a {\em merging}-BIC, and it is surrounded by resonant
modes of ultra-high $Q$ factor, and thus it is also a {\em super}-BIC. Our
theory reveals that a non-generic BIC is in fact a {\em merging}-BIC for
$\delta \to 0$ and almost any perturbation profile $F$.

The rest of this paper is organized as follows. 
In Section 2, we recall some facts about resonant modes, leaky modes,
and BICs in waveguides with lateral leakage channels,  and introduce
generic and non-generic BICs.
In Section 3, we analyze resonant and leaky modes near a BIC, in a
fixed waveguide, using a perturbation method.  
In Section 4, a bifurcation theory for BICs in a perturbed waveguide
is developed based on power series in $\sqrt{\delta}$. 
To illustrate our theory, numerical examples are presented in Sections
3 and 4. The paper is concluded with some remarks in Section 5.

\section{Basic definitions}

\begin{figure}[thbp] 
	\centering 
	\includegraphics[width=7cm]{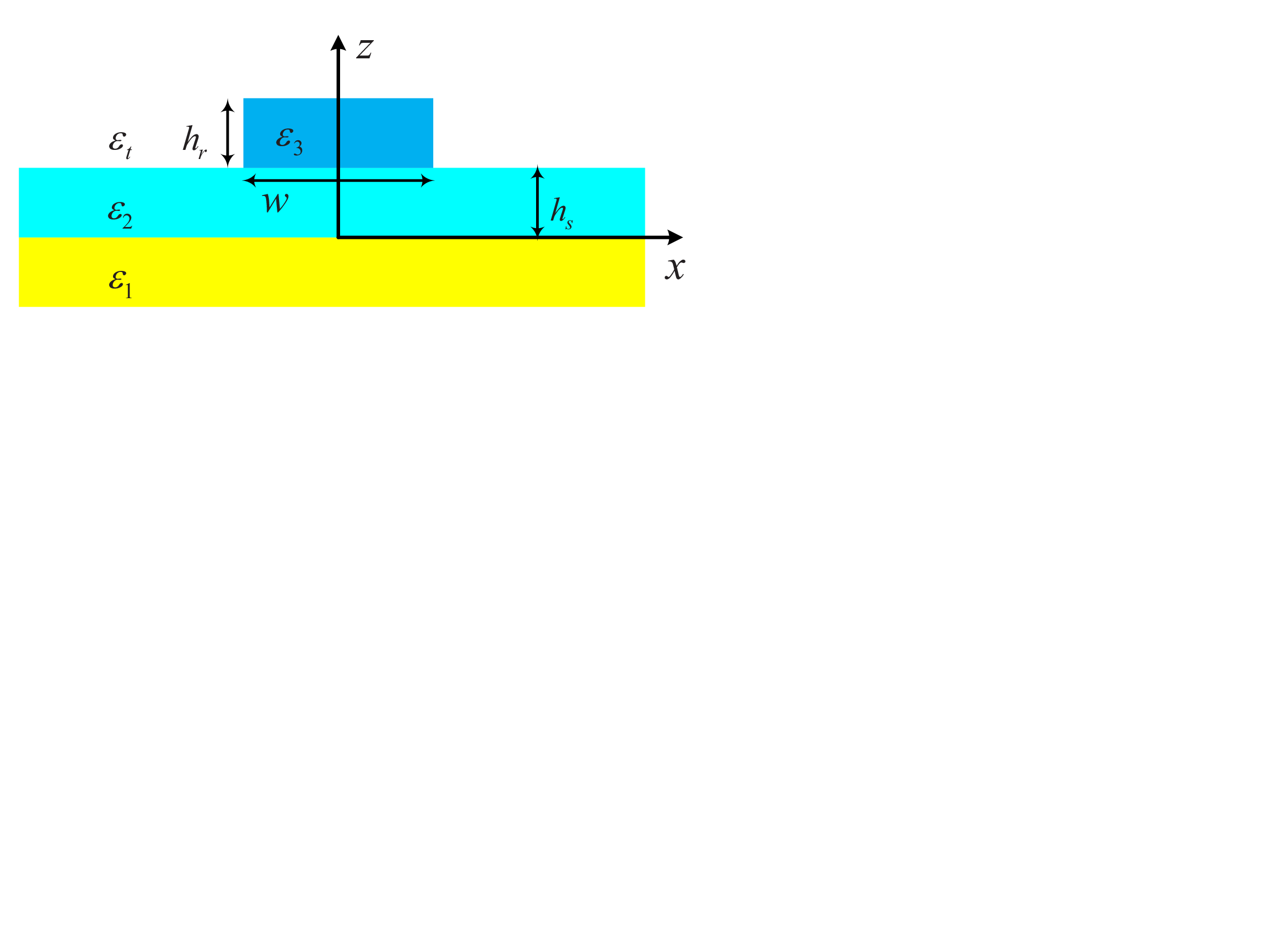}
	\caption{A ridge waveguide with a rectangular core of width
		$w$ and height $h_r$. The background is a slab waveguide with a slab of
		thickness $h_s$. The dielectric constants of the
		substrate (yellow region), the slab (light cyan
		region), the core (light blue region) and the cladding are $\varepsilon_1$, $\varepsilon_2$,
		$\varepsilon_3$ and $\varepsilon_t$, respectively.}
	\label{ridgewg}
\end{figure}

We consider a three-dimensional (3D) $y$-invariant lossless open
dielectric waveguide consisting of a waveguide core and a layered
background which itself is a planar waveguide parallel to the $xy$ plane, 
where $y$ is the waveguide axis and $x$ is the lateral variable of the 3D waveguide.
The dielectric function of the structure depends only on
two transverse variables $x$ and $z$, i.e., $\varepsilon=\varepsilon(x,z)$.
The dielectric function $\varepsilon_b$ of the layered background
depends only on $z$. The waveguide core occupies a
bounded domain in the $xz$ plane. We further assume that $\varepsilon$
is symmetric about $x$, i.e., $\varepsilon(x,z)=\varepsilon(-x,z)$. 
As an example, we show a ridge waveguide in Fig.~\ref{ridgewg}. 

For a guided mode propagating along the $y$ axis, the electric field
can be written as  $\mbox{Re} [ \bE(\br) e^{-i\omega t}]$,  
where $\omega$ is the angular frequency, $\br=(x,z)$, 
$\bE=\bp(\br)e^{i\beta y}$, $\beta$ is the propagation constant,
and $\bp(\br) \to {\bf 0}$ as $|\br|=\sqrt{x^2+z^2} \to \infty$. 
The frequency-domain Maxwell's equations give rise to the following
equation for the complex amplitude $\bp$:
\begin{equation}
	\label{reducewaveeq}
	(\nabla+i\beta \2e)\times(\nabla+i\beta \2e)\times
	\bp-k^2\varepsilon(\br)\bp=0, 
\end{equation}
where $k=\omega/c$ is the free space wavenumber and $\hat{y}$ is the
unit vector in the $y$ direction. Since the field must decay as $z \to
\pm \infty$, the propagation constant satisfies 
$\beta > k \max \{ \sqrt{\varepsilon_1}, \sqrt{\varepsilon_t} \}$.
For the same frequency, the background planar waveguide may support
a few guided modes. We order the eigenmodes of the planar waveguide according to
their propagation constants, denote the propagation constant of the
$j$-th transverse electric (TE) mode by $\eta_j^{\rm te}$ and the
corresponding vertical profile by $u_j(z)$, and those of the $j$-the 
transverse magnetic (TM) mode by 
$ \eta_j^{\mathrm{tm}}$ and $v_j(z)$. Both $u_j(z)$ and $v_j(z)$ are
real functions and they can be normalized so that 
\begin{equation}
	\label{tetmmn}
	\frac{1}{L} \int_{-\infty}^\infty |u_j(z)|^2\rmd z=1,\quad
	\frac{1}{L} \int_{-\infty}^\infty \frac{1}{\varepsilon_b(z)}|v_j(z)|^2\rmd z=1,
\end{equation}
where $L$ is a characteristic length. Typically, the propagation constants satisfy
\begin{equation}\label{key}
	\eta_1^{\mathrm{te}}>\eta_1^{\mathrm{tm}}>\eta_2^{\mathrm{te}}>\eta_2^{\mathrm{tm}}>\cdots. 
\end{equation}
Thus, the first TE mode has the largest propagation constant, i.e.,
$\eta_{\rm max} = \eta_1^{\rm te}$. 

If $\beta>\eta_{\rm max}$, the guided mode is a classical one and it
depends on $\beta$ and 
$\omega$ continuously. A BIC is a special guided mode with 
$\beta<\eta_{\rm max}$, and it corresponds to an isolated
point in the $\beta$-$\omega$ plane. In this paper, we focus on BICs
with $\beta$ satisfying $\eta_1^{\mathrm{tm}} < \beta <
\eta_1^{\mathrm{te}}$.  In that case, the BIC is 
compatible with the left- and
right-going first TE mode  ${\bf u}^\pm e^{ i (\beta y - \omega t)}$, 
where 
\begin{equation}\label{u}
	{\bf u}^{\pm}=\frac{i}{\eta_1^{\mathrm{te}}}\left[
	\begin{array}{c}
		\mp\beta\\
		\alpha_1^{\mathrm{te}}\\
		0\\
	\end{array}
	\right]u_1(z)e^{\pm i\alpha_1^{\mathrm{te}} x},\quad
	\alpha_1^{\mathrm{te}}=\sqrt{\left(\eta_1^{\mathrm{te}}\right)^2-\beta^2}
	> 0. 
\end{equation}
Since the BIC is a guided mode, it must decay as $x \to \pm \infty$
and cannot couple with ${\bf u}^+$ or ${\bf u}^-$. Clearly, we can
formulate a scattering problem by sending right-going incident wave
$C^+ {\bf   u}^+$ from $x=-\infty$ and left-going incident wave $C^- {\bf u}^-$
from $x=+\infty$, where $C^+$ and $C^-$ are given constants. 
The incident waves give rise to outgoing waves $D^- {\bf u}^-$ and
$D^+ {\bf u}^+$ for $x \to -\infty$ and $x \to +\infty$,
respectively. Because of the BIC, the solution of the scattering
problem is not unique, but the amplitudes of the outgoing waves $D^+$ and
$D^-$ are well-defined and related to $C^+$ and $C^-$ by a  $2 \times
2$ scattering matrix. Since $\beta > k \max \{ \sqrt{ \varepsilon_1},
\sqrt{ \varepsilon_t} \}$, the incident waves will not induce
outgoing waves in the substrate and the cladding. Therefore, power  is
balanced, the scattering matrix is unitary, and $|C^+|^2 + |C^-|^2 =
|D^+|^2 + |D^-|^2$. 

Since the structure is lossless and symmetric in $x$, the BIC and the
corresponding scattering solutions can be scaled to have some useful
symmetry.  Let $\P$ and $\T$ be operators satisfying 
\begin{equation}\label{key}
	\begin{aligned}
		{\mathcal{P}}{\bf f}=\left[
		\begin{array}{c}
			\displaystyle	-f_x(-x,z)\\
			f_y(-x,z)\\
			f_z(-x,z)\\
		\end{array}
		\right],\quad {\mathcal{T}}{\bf f}=\left[
		\begin{array}{c}
			\displaystyle	\overline{f}_x(x,z)\\
			\displaystyle -\overline{f}_y(x,z)\\
			\overline{f}_z(x,z)\\
		\end{array}
		\right],
	\end{aligned}
\end{equation}
where ${\bf f} = {\bf f}(x,z)$ is an arbitrary vector function and
$\overline{f}_x$ is the complex conjugate of $f_x$. 
If the BIC $\{ k,\beta,\bp \}$ is 
non-degenerate, we have either $\P\bp=\bp$ or $\P\bp=-\bp$, and it can
be scaled such that $\T\bp=\bp$. For the same $k$ and $\beta$ as the
BIC, by choosing $C^- = \pm C^+$, we can construct two scattering
solutions satisfying $\P\bs=\pm\bs$, where $\bs$ is the complex amplitude of the electric
field. Moreover, the
scattering solutions can be further scaled and shifted such that 
$\T\bs=\bs$ and $\left<\varepsilon\bs,\bp\right>=0$, 
where $\left< \cdot , \cdot \right>$ is the inner product defined as
\[
\left< {\bf u}, {\bf v} \right> = \frac{1}{L^2} \int_{\mathbb{R}^2} \overline{\bf
	u} \cdot {\bf v} \, d{\bf r}.
\]

We are concerned with non-generic
BICs satisfying the following condition
\begin{equation}
	\label{ngBICdef}
	\left<{\bf \Psi}, \B \bp \right> = 0, 
\end{equation}
where $\bs$ is the one with the same parity symmetry (i.e. operation
by $\P$) as the BIC, and 
$\B$ is the operator satisfying 
\begin{equation}
	\label{defopB}
	\B{\bf w}=-i\left[\left(\nabla+i\beta
	\2e\right)\times\2e+\2e\times\left(\nabla+i\beta \2e\right)\right]\times{\bf w}
\end{equation}
for any differentiable vector function ${\bf w}(x,z)$.
Condition (\ref{ngBICdef}) was identified in the robustness theory for BICs
in waveguides with lateral leakage channels~\cite{Yuan21OE}. 
It has been
proved that if  the BIC is generic, i.e., Eq.~(\ref{ngBICdef}) is not
satisfied, and $\eta_1^{\rm tm} < \beta < \eta^{\rm te}_1$, then
it is robust with respect to any small perturbation that preserves the
lateral mirror symmetry~\cite{Yuan21OE}.

Given a particular BIC with frequency $\omega_*$ and propagation
constant $\beta_*$, we can consider resonant and leaky modes for
$\beta$ near $\beta_*$ and $\omega$ near $\omega_*$,
respectively. Both resonant and leaky modes satisfy outgoing radiation
conditions as $x \to \pm \infty$, and they are coupled with outgoing
first TE mode  ${\bf u}^\pm$.  In other words,
the complex electric-field amplitude $\bp$ of a resonant or leaky mode
has the following asymptotic relation
\begin{equation}
	\label{farfield}
	\bp ({\bf r}) \sim c_{1,{\rm te}}^\pm {\bf u}^\pm, \quad x\to \pm
	\infty, 
\end{equation}
where $c_{1,{\rm te}}^\pm$ are nonzero coefficients. A resonant mode
is defined for a real $\beta$. Since power is radiated out laterally to $x = \pm
\infty$, the amplitude of the resonant mode must decay with time,
thus, $\omega$ is complex and $\mbox{Im}(\omega) < 0$. As a result, 
the TE and TM modes of the background planar waveguides are eigenmodes
of 1D Helmholtz equations with a complex freespace wavenumber
$k$. All propagation constants $\eta_j^{\rm te}$ and  $\eta_j^{\rm
	tm}$ have a negative imaginary part. Therefore, $\mbox{Im}(
\alpha_1^{\rm te}) < 0$ and ${\bf u}^{\pm}$ diverges as $x \to \pm
\infty$. A leaky mode is defined for a real frequency $\omega$. It
also loses power laterally, and must decay as it propagates
forward. This implies that $\beta$ is complex and $\mbox{Im}(\beta) >
0$. Since the frequency is real, the propagation constants of the
background planar waveguide are real, but since $\beta$ is complex, we
still have a complex $\alpha_1^{\rm te}$ with a negative imaginary
part, and ${\bf u}^\pm$ also diverges as $x \to \pm \infty$. 

\section{Resonant and leaky modes near BICs}

In this section, we use a perturbation method to analyze the resonant
and leaky modes near a BIC in an optical waveguide described in the beginning of Section 2. 
We consider a BIC with a real frequency $\omega_*$ (freespace
wavenumber $k_* = \omega_*/c$), a real propagation constant $\beta_*$,
and a complex electric-field amplitude $\bp_*$. 
Without loss of generality, we assume $\P\bp_*=\bp_*$.
The case for $\P\bp_*=-\bp_*$ is similar. 
We further scale and normalize the BIC such that
$\T\bp_*=\bp_*$ and $\left<\varepsilon\bp_*,\bp_*\right>=1$.
The scattering solution can be chosen to satisfy
\begin{equation}\label{Conditionsca}
	\P\bs_*=\bs_*,\;\T\bs_*=\bs_*,\;\left<\varepsilon\bs_*,\bp_*\right>=0.
\end{equation}

We are concerned with resonant and leaky modes for $\beta$ near
$\beta_*$ and $\omega$ near $\omega_*$, respectively.
Our theory reveals a major distinction between the generic and non-generic BICs.
Near a generic BIC, $\mbox{Im}(\omega)$ of the resonant modes is
proportional to $|\beta-\beta_*|^2$, and $\mbox{Im}(\beta)$ of the
leaky modes is proportional to $|\omega -\omega_*|^2$. But near a
non-generic BIC, the imaginary parts of $\omega$ and $\beta$
of the resonant and leaky modes are much
smaller, and they typically exhibit a fourth order dependence
on $|\beta-\beta_*|$ and $|\omega - \omega_*|$, respectively. 

\subsection{Resonant modes: perturbation with respect to $\beta$}

We first analyze the resonant modes near a BIC.
For any real $\beta$ near $\beta_*$, there is a resonant mode near
the BIC. If $\delta=(\beta-\beta_*)L$ is small, we can expand the freespace
wavenumber $k=\omega/c$ and 
complex electric-field amplitude $\bp$ of the resonant mode 
in power series of $\delta$: 
\begin{align}
	\label{Ask}
	k  &=k_*+\delta k_1+\delta^2 k_2+\delta^3 k_3+\delta^4k_4+\cdots,\\
	\label{AsE}
	\bp &=\bp_*+\delta\bp_1+\delta^2\bp_2 +\delta^3\bp_3 +\delta^4\bp_4+\cdots.
\end{align}
Our objective is to determine the leading order for the imaginary part of $k$.
We show that if the BIC is generic, then $\mbox{Im}(k)\sim\delta^2\mbox{Im}(k_2)$ and $\mbox{Im}(k_2)< 0$;
if the BIC is non-generic, then $\mbox{Im}(k_2)= 0$ and typically $\mbox{Im}(k)\sim\delta^4\mbox{Im}(k_4)$ 
with a negative $\mbox{Im}(k_4)$. 

To obtain the above results, we substitute Eqs.~(\ref{Ask})-(\ref{AsE}) into
Eq.~(\ref{reducewaveeq}), collect the $O(1)$ terms, and obtain
the following equation satisfied by the BIC:
\begin{equation}
	\L\bp_*:=(\nabla+i\beta_*\2e)\times(\nabla+i\beta_*\2e)\times\bp_*-k_*^2\varepsilon\bp_*=0.
\end{equation}
The above equation defines an operator $\L$ and  it satisfies $\L\T=\T\L$ and $\L\P=\P\L$. 
Collecting the $O(\delta^j)$ terms, we obtain 
\begin{align}
	\label{Lphi1Equ}
	\L\bp_1&={\bf R}_1(\bp_*;k_1):=\B\bp_*/L+2k_*k_1\varepsilon\bp_*,\\
	\label{LphijEqu}
	\L\bp_j&={\bf R}_j(\bp_*;\bp_1,\cdots,\bp_{j-1};k_1,\cdots,k_{j}),\;j\geq 2,
\end{align}
where $\B$ is the operator defined in Eq.~(\ref{defopB}) with $\beta$
replaced by $\beta_*$. 
The right hand sides ${\bf R}_j$ are listed in Appendix A. 
{As shown in Refs. \cite{Yuan21OE,Yuan20PRAPert}, a differential equation $\L{\bf w}={\bf f}$ is solvable if and only if $\left<{\bp_*},{\bf f}\right>=0$.
	If $\P{\bf f}={\bf f}$ and ${\bf f}\rightarrow 0$ as $|\br|\rightarrow\infty$, there exists a particular solution $\bf w$ that satisfies $\P{\bf w}=\bf w$ and has asymptotic behavior
	${\bf w}\sim d{\bf u}_*^\pm$ as $x\rightarrow \pm \infty$, where ${\bf u}_*^{\pm}$ is defined as in Eq.~(\ref{u}) with
	$\beta$ replaced by $\beta_*$, $k$ replaced by $k_*$, etc. 
	Moreover, the coefficient $d$ is a multiple of the integral $\left<{\bs_*},{\bf f}\right>$. If $\left<{\bs_*},{\bf f}\right>=0$, we have $d=0$ and then ${\bf w}\rightarrow 0$ as $|\br|\rightarrow\infty$.}

The solvability condition of Eq.~(\ref{Lphi1Equ}), i.e., $\left<\bp_*,{\bf R}_1\right>=0$, leads to $2k_*k_1=-\left<\bp_*,\B\bp_*\right>/L$.
Moreover, since $\T\B=\B\T$, we can show that $k_1$ is real.
{With $k_1$ determined, 
	we have $\P{\bf R}_1={\bf R}_1$ and ${\bf R}_1\rightarrow 0$ as $|\br|\rightarrow\infty$.}
Equation~(\ref{Lphi1Equ}) has a particular solution $\bp_1$ that
satisfies $\P\bp_1 =\bp_1$ and has the following asymptotic form
\begin{equation}\label{key}
	\bp_1\sim d_1 {\bf u}_*^{\pm}, \quad x\to \pm \infty, 
\end{equation}
where $d_1$ is a constant and a multiple of $\left<{{\bf \Psi}_*}, {\bf R}_1 \right>$.

A formula for $k_2$ can be deduced from the solvability condition of
Eq.~(\ref{LphijEqu}) with $j=2$.
As shown in Appendix A, this condition implies that the imaginary part of
$k_2$ is proportional to $-|d_1|^2$.
Since the amplitude $\bs_*$ is chosen to satisfy the Eq.~(\ref{Conditionsca}), we have 
$\left<{{\bf \Psi}_*}, {\bf R}_1 \right>=\left<{{\bf \Psi}_*}, \B \bp_* \right>/L$.
Therefore, if the BIC is generic, i.e., $\left<{{\bf \Psi}_*}, \B \bp_* \right> \neq 0$, we have $d_1\ne 0$, $\mbox{Im}(k_2) < 0$, $\mbox{Im}(\omega) \sim |\beta-\beta_*|^2$,  and $Q\sim |\beta-\beta_*|^{-2}$.

On the other hand, if the BIC is non-generic, we have $\left<{{\bf \Psi}_*}, \B \bp_* \right> = 0$, thus $d_1=0$ and $\mbox{Im} (k_2)=0$.
Moreover, we must have $\mbox{Im}( k_3)=0$, since otherwise $\mbox{Im}(\omega)$ will change sign as $\beta$ passes
through $\beta_*$. This is not possible, because any resonant mode
with radiation loss  must
have $\mbox{Im}(\omega) < 0$, so that the field amplitude can 
decay with time. 
With $k_1$, $k_2$, and $\bp_1$ determined, as shown in Appendix A, Eq.~(\ref{LphijEqu}) with $j=2$ has a particular solution $\bp_2$ which satisfies $\P\bp_2 =\bp_2$ and has the following asymptotic form
\begin{equation}\label{d2}
	\bp_2 \sim d_2 {\bf u}_*^{\pm}, \quad x \to \pm \infty,
\end{equation}
where $d_2$ is the coefficient. 
Moreover, we show that the imaginary part of $k_4$ is proportional to $-|d_2|^2$ in Appendix A.
Therefore, if $d_2 \ne 0$, we have $\mbox{Im}(\omega) \sim
|\beta-\beta_*|^4$ and  $Q\sim |\beta-\beta_*|^{-4}$ for a non-generic BIC. 
Consequently, the resonant mode near a non-generic BIC has an ultra-high $Q$ factor.

\subsection{Leaky modes: perturbation with respect to $\omega$}

Next, we analyze the leaky modes near a BIC. For any real $\omega$ near $\omega_*$, the waveguide supports a leaky mode with
a complex propagation constant $\beta$ and complex electric-field amplitude $\bp$.
If $\delta=(k-k_*)L$ is small, we can expand the propagation constant
$\beta$ and  complex electric-field amplitude $\bp$ of the leaky mode 
in power series of $\delta$: 
\begin{align}
	\label{Asg}
	\beta&=\beta_*+\delta\beta_1+\delta^2\beta_2+\delta^3\beta_3+\delta^4\beta_4\cdots,\\
	\label{AsE2}
	\bp&=\bp_*+\delta\bp_1+\delta^2\bp_2+\delta^3\bp_3+\delta^4\bp_4\cdots.
\end{align}
Substituting Eqs.~(\ref{Asg})-(\ref{AsE2}) into
Eq.~(\ref{reducewaveeq}) and collecting $O(\delta^j)$ terms, we obtain 
\begin{align}
	\label{Lphi1Equleaky}
	\L\bp_1&={\bf L}_1(\bp_*;\beta_1):=\beta_1\B\bp_*+2k_*\varepsilon\bp_*/L,\\
	\label{LphijEquleaky}
	\L\bp_j&={\bf L}_j(\bp_*;\bp_1,\cdots,\bp_{j-1};\beta_1,\cdots,\beta_{j}),\;j\geq 2.
\end{align}
where ${\bf L}_j$ are listed in Appendix A. 

The solvability condition of Eq.~(\ref{Lphi1Equleaky}) leads to a real $\beta_1L=-2k_*/\left<\bp_*,\B\bp_*\right>$.
The integral $\left<\bp_*,\B\bp_*\right>$ is typically non-zero.
With $\beta_1$ determined, following the same process as in the previous subsection, we can show that if 
the BIC is generic, then $\mbox{Im}(\beta_2)> 0$ and $\mbox{Im}(\beta)\sim |\omega-\omega_*|^2$.
On the other hand, if the BIC is non-generic, 
then $\mbox{Im}(\beta_2) = \mbox{Im}(\beta_3)=0$, and typically $\mbox{Im}(\beta_4)>0$.
In that case, the leaky mode near a non-generic BIC has $\mbox{Im}(\beta)\sim |\omega-\omega_*|^4$. 
Consequently, a leaky mode near a non-generic BIC has ultra-low
leakage loss.

\subsection{Numerical examples}

\begin{figure}[h]
	\centering
	\includegraphics[width=6.5cm]{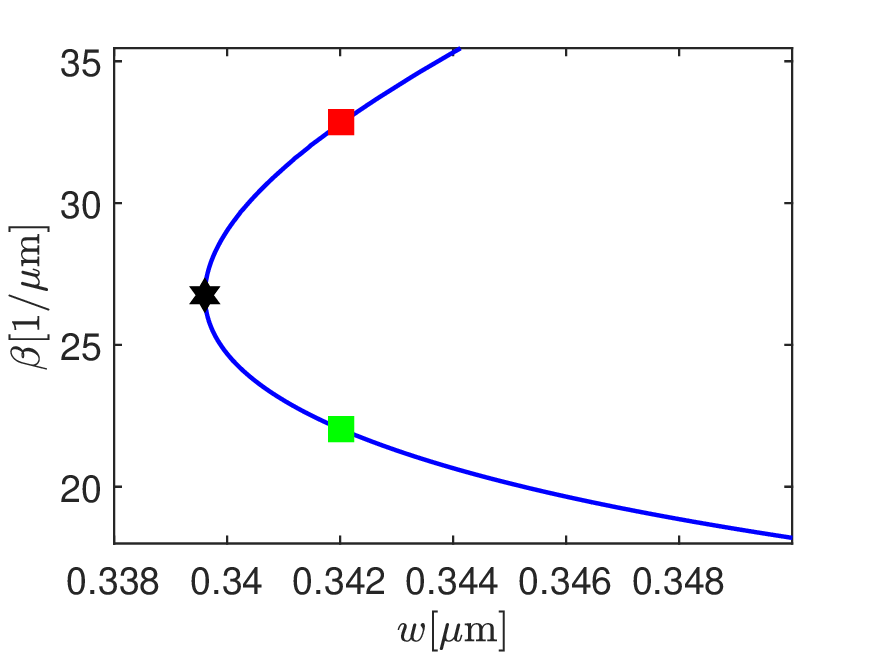}\;
	\includegraphics[width=6.5cm]{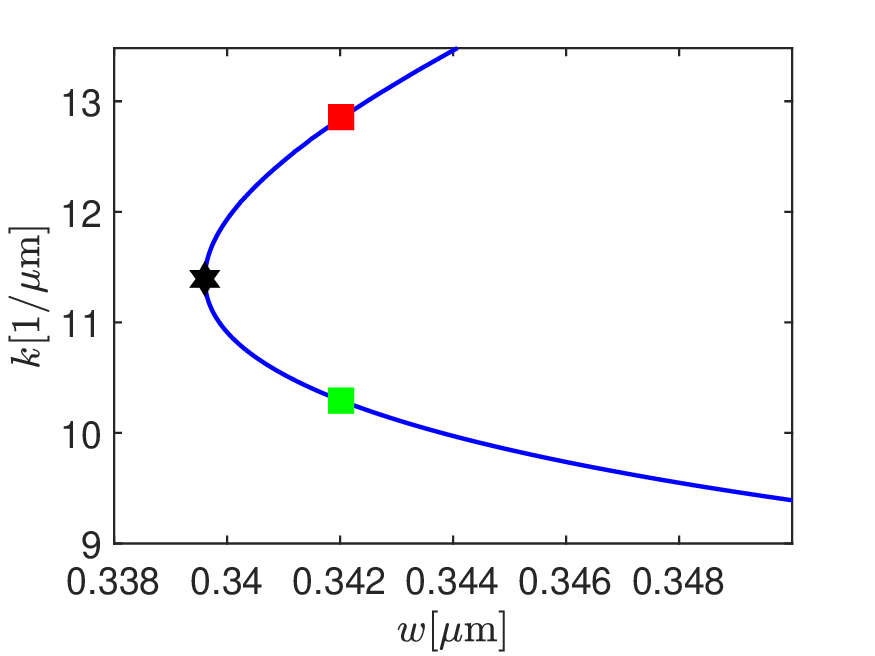} 
	\caption{The wavenumber $k$ and propagation constant $\beta$ of BICs for different width $w$.}\label{betakBIC}
\end{figure}

\begin{figure}[h]
	\centering
	\includegraphics[width=6.5cm]{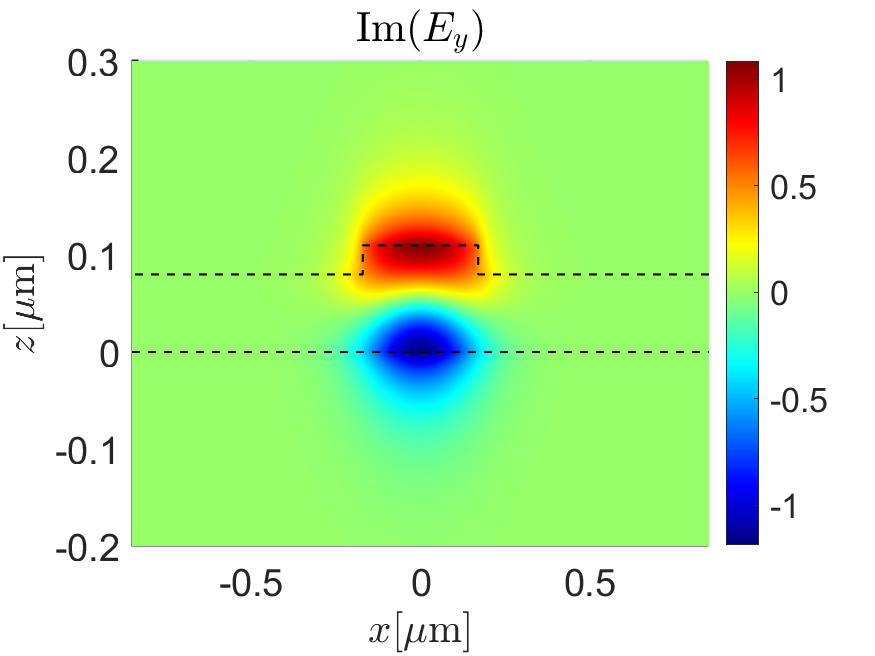}\;
	\includegraphics[width=6.5cm]{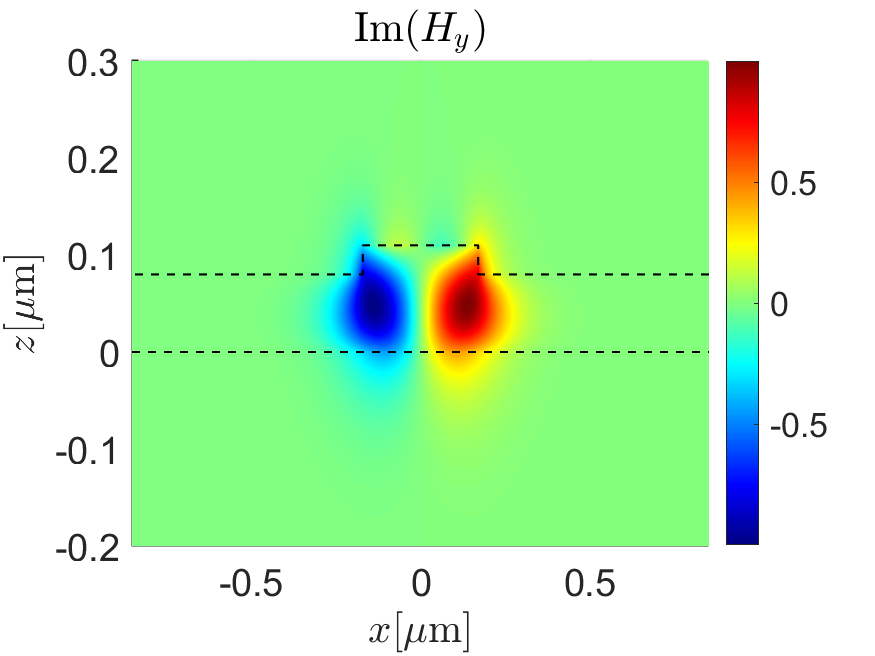}
	\caption{The imaginary parts of $E_y$ and $H_y$ for the
		non-generic BIC at $w=w_\natural$.}\label{field}
\end{figure}

\begin{figure}[h]
	\centering
	\includegraphics[width=6.5cm]{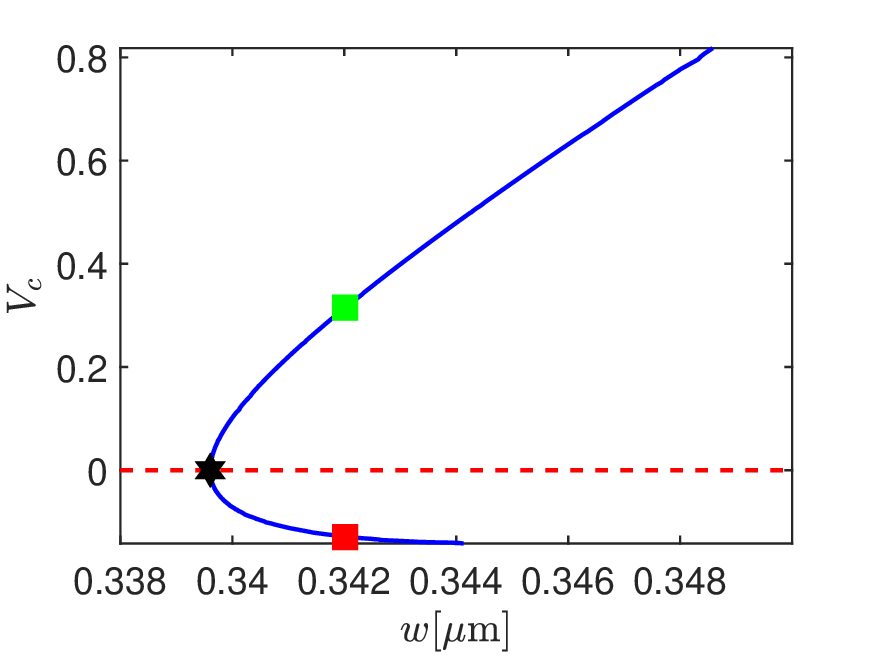}
	\caption{The quantity $V_c$ for different BICs shown in Fig.~\ref{betakBIC}.}\label{Vc}
\end{figure}

To validate our theory, we consider a silicon rib waveguide 
with silica substrate and air cladding, as shown in Fig. \ref{ridgewg}.
The dielectric constants are $\varepsilon_t=1$, $\varepsilon_1=2.1025$, 
and $\varepsilon_2=\varepsilon_3=11.0304$.
The height of the ridge and the thickness of the slab are $h_r=$ 0.03 $\mum$ and $h_s=$  0.08 $\mum$, respectively.
We consider a non-degenerate BIC satisfying $\P\bp_*=\bp_*$.
By tuning the width of the ridge, a {\em merging}-BIC is obtained at $w=w_\natural\approx 0.3396$ $\mum$.  
The wavenumber $k$ and propagation constant $\beta$ of the BICs for different width $w$ are shown in Fig. \ref{betakBIC}. 
The {\em merging}-BIC is marked by a black hexagon. 
The imaginary part of electromagnetic field components $E_y$ and $H_y$ of the {\em merging}-BIC are shown in Fig. \ref{field}.
In Fig. \ref{Vc}, we show the quantity $V_c=\left<{{\bf \Psi}_*}, \B \bp_* \right>$ for different BICs.
It is clear that for the {\em merging}-BIC at $w=w_\natural$, we have
$V_c=0$. Therefore, the {\em merging}-BIC is indeed a non-generic BIC.

\begin{figure}[h]
	\centering
	\includegraphics[width=6.5cm]{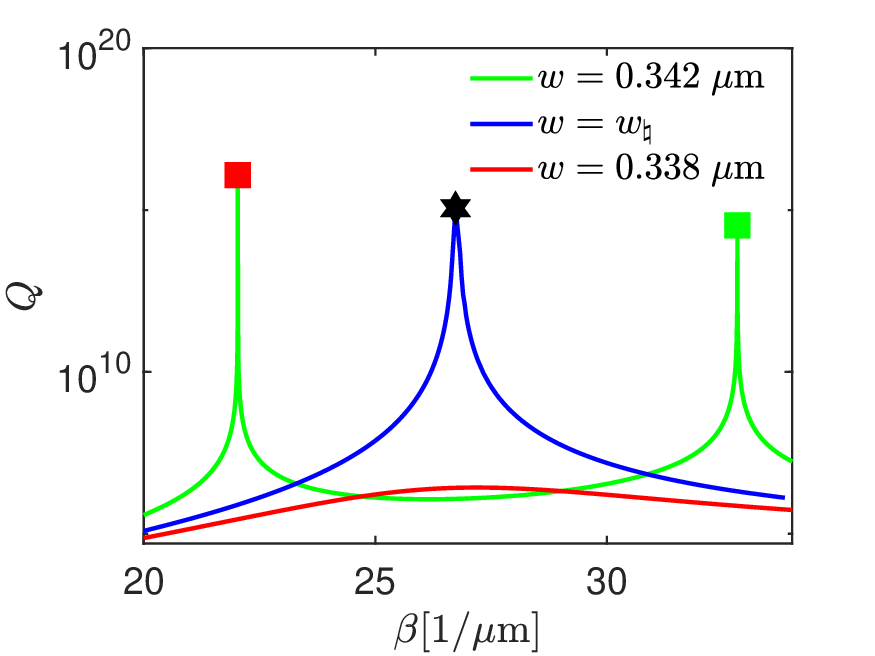}\;
	\includegraphics[width=6.5cm]{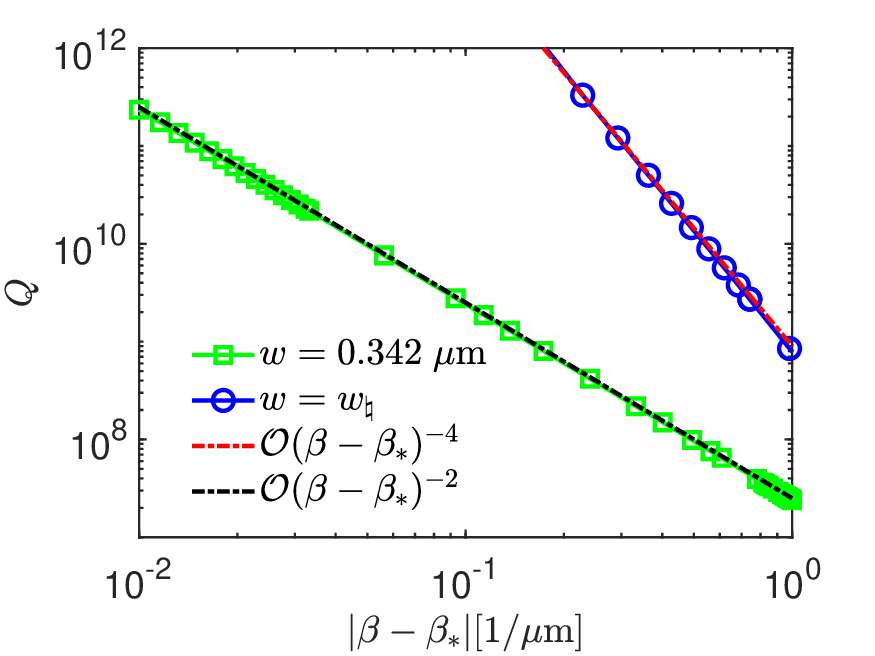}
	\caption{The $Q$ factor of resonant modes for three different
		values of width $w$. In the right panel,
		$\beta_*\approx32.8168[1/\mum]$ for
		$w=0.342\;\mum$.}\label{resonance} 
\end{figure}
In Fig. \ref{resonance}, we show the $Q$ factor of resonant modes for three different values of $w$. 
For $w=0.342\;\mum$, the waveguide has two BICs corresponding to the red and green squares in Figs. \ref{betakBIC}, \ref{Vc}, and \ref{resonance}.
As shown in Fig. \ref{resonance}, the $Q$ factor of the resonant modes near these two BICs satisfies $Q\sim|\beta-\beta_*|^{-2}$.
For $w=w_\natural$, there is only one non-generic BIC and the $Q$ factor satisfies $Q\sim|\beta-\beta_*|^{-4}$.
As shown in Fig. \ref{resonance}, for $w=0.338\;\mum<w_\natural$, there is no BIC in the waveguide, and
there are only resonant modes with a finite $Q$ factor.

\begin{figure}[h]
	\centering
	\includegraphics[width=6.5cm]{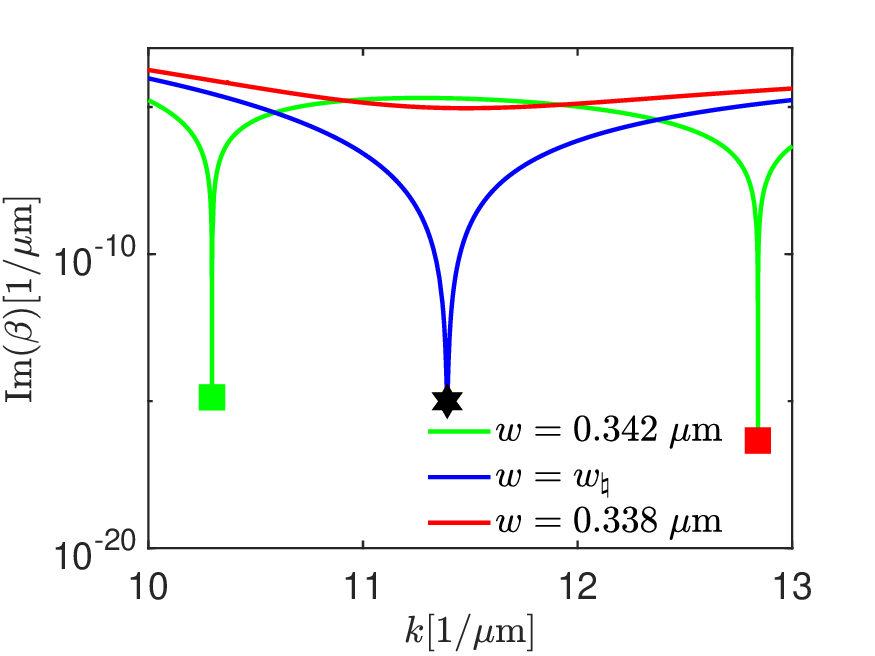}\;
	\includegraphics[width=6.5cm]{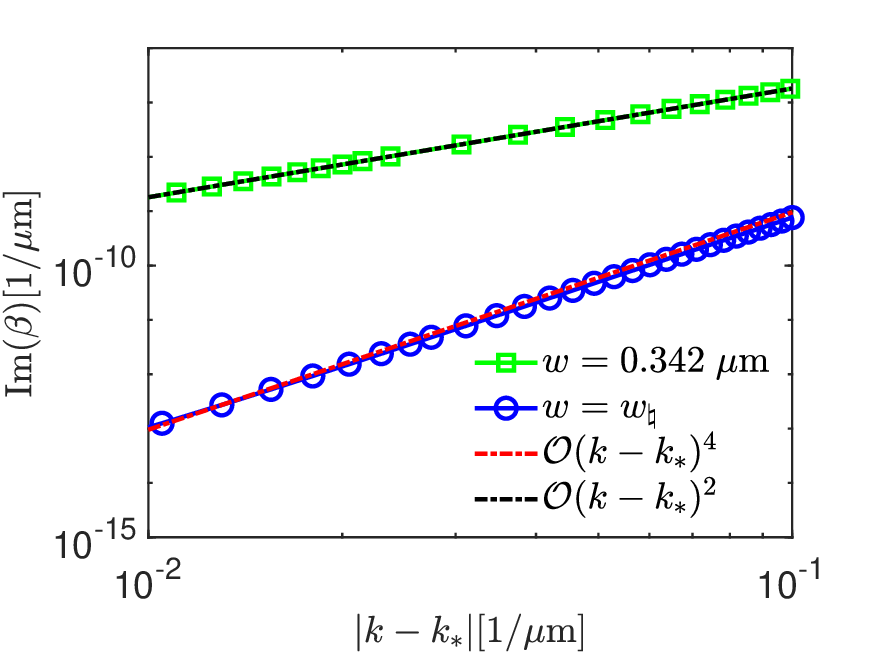}
	\caption{$\mbox{Im}(\beta)$ of leaky modes for three different
		values of width $w$. In the right panel, $k_*\approx12.8403[1/\mum]$ for $w=0.342\;\mum$.}\label{leakage}
\end{figure}

In Fig. \ref{leakage}, 
we show the imaginary part of $\beta$ of leaky modes for three different values of $w$. 
For $w=0.342\;\mum$, it is clear that $\mbox{Im}(\beta)$ of the leaky modes near the two BICs satisfies 
$\mbox{Im}(\beta)\sim|\omega-\omega_*|^{2}$.
For $w=w_\natural$, $\mbox{Im}(\beta)$ satisfies $\mbox{Im}(\beta)\sim|\omega-\omega_*|^{4}$.
For $w=0.338\;\mum<w_\natural$, the waveguide can only support leaky modes.

\section{Bifurcation theory for non-generic BICs}

In the previous section, we found a {\em merging}-BIC by tuning the ridge width, and showed that the {\em merging}-BIC is in fact a non-generic BIC.
We also showed that the waveguide has two BICs for $w>w_\natural$ and no BIC for $w<w_\natural$.
Notice that a small change of $w$ around $w_\natural$ can be regarded as a perturbation of the waveguide.
In this section, we consider a general perturbation to waveguides with a non-generic BIC, 
and analyze the existence of BICs in the perturbed waveguide.

Using the same notations for the unperturbed waveguide and the non-generic BIC, we consider a perturbed waveguide with
a dielectric function given by
\begin{equation}\label{perdielec}
	\varepsilon(\br)=\varepsilon_*(\br)+\delta F(\br),
\end{equation}
where $\delta$ is a small real number and $F(\br)$ is a real function 
of $x$ and $z$. We further assume that $F$ is symmetric in $x$ and has compact support. 
In the previous work on robustness \cite{Yuan21OE}, 
BICs in the perturbed waveguide are constructed through
power series of $\delta$  
by using the condition $\left<\bs_*,\B\bp_*\right>\ne 0$. 
Therefore, this robustness theory is not applicable to non-generic BICs 
satisfying $\left<\bs_*,\B\bp_*\right>=0$.
In the following, we assume the non-generic BIC in the unperturbed waveguide has a non-zero $d_2$ [defined in Eq. (\ref{d2})], 
and introduce a characteristic function $\chi(F)$ given by
\begin{equation}
	\chi(F)=-k_*^2\left<F{{\bf \Psi}_*},\bp_*\right>/A,
\end{equation}
where $A$ is proportional to $d_2$ and independent of $F$.
It can be proved that $\chi(F)$ is real, and it is clear that $\chi(-F)=-\chi(F)$.
Our main result is that for a sufficiently small $\delta$, if $\chi(F)>0$, then the perturbed waveguide has two
BICs for $\delta>0$ and no BIC for $\delta<0$, and 
if $\chi(F)<0$, then the perturbed waveguide has two
BICs for $\delta<0$ and no BIC for $\delta>0$.

In the remainder of this section, we focus on the case $\delta>0$ and $\chi(F)>0$, and show that
there indeed exist two BICs which are given by power series of $\sqrt{\delta}$:
\begin{equation}
	\label{Asthree}
	k=k_*+\sum_{j=1}^\infty k_j\delta^{j/2},\;\beta=\beta_* +\sum_{j=1}^\infty \beta_j\delta^{j/2},\;
	\bp=\bp_*+\sum_{j=1}^\infty \bp_j\delta^{j/2},
\end{equation}
where $k$, $\beta$, and $\bp$ are the freespace wavenumber, the propagation constant, and the complex electric-field amplitude of these two BICs, respectively.
To justify the existence of these BICs, we need to show for each
$j\geq 1$, $k_j$ and $\beta_j$ can be solved and they are real, $\bp_j$ decays rapidly to zero as $x\rightarrow \pm\infty$, and it can be chosen to satisfy
\begin{equation}\label{symort}
	\P\bp_j=\bp_j,\;\T\bp_j=\bp_j,\;\left<\varepsilon_*\bp_j,\bp_*\right>=0.
\end{equation}
In addition, there are two solutions for $k_j$, $\beta_j$ and $\bp_j$ corresponding to the two BICs.

To prove the above results, we first substitute Eq. (\ref{Asthree}) into Eq. (\ref{reducewaveeq}), collect terms of different powers of $\delta^{j/2}$, and obtain
\begin{align}
	\label{j1}
	\L\bp_1&={\bf B}_1(\bp_*;\beta_1,k_1):=\beta_1\B\bp_*+2k_*k_1\varepsilon_*\bp_*,\\
	\label{jj}
	\L\bp_j&={\bf B}_j(\bp_*;\bp_1,\cdots,\bp_{j-1};\beta_1,k_1,\cdots,\beta_{j},k_{j}),\;j\geq 2,
\end{align}
where the right hand sides ${\bf B}_j$ are listed in Appendix B. 
{For the equation of $\bp_j$ to have a solution that decays rapidly to zero as $x\rightarrow\pm\infty$, the right hand side ${\bf B}_j$ must satisfy the following two conditions} 
\begin{equation}
	\left<{\bp}_*,{\bf B}_j\right>= 0,\;\left<{{\bf \Psi}_*},{\bf B}_j\right>= 0.
\end{equation}

Since the original BIC is non-generic and $\bs_*$ is chosen to satisfy $\left<\varepsilon_*\bs_*,\bp_*\right>=0$, we obtain $\left<{{\bf \Psi}_*},{\bf B}_1\right>= 0$.
The condition $\left<{\bp}_*,{\bf B}_1\right>= 0$ leads to a real linear relation $2k_*k_1=-\beta_1\left<\bp_*,\B\bp_*\right>$.
Using this result, as shown in Appendix B, the condition $\left<{{\bf \Psi}_*},{\bf B}_2\right>=0$ 
gives rise to a real quadratic equation of $\beta_1$:
\begin{equation}
	\beta_1^2A+k_*^2\left<F{{\bf \Psi}_*},\bp_*\right>=0,
\end{equation}
where $A$ is mentioned earlier in this section.
Since $\chi(F)>0$, we obtain two real $\beta_1$ given by
\begin{equation}\label{beta1}
	\beta_1=\pm\sqrt{\chi(F)}.
\end{equation}
For each $\beta_1$, we have a real $k_1$ and Eq.~(\ref{j1}) has a particular solution $\bp_1$ 
that satisfies Eq.~(\ref{symort}) and decays to zero as $x\rightarrow\pm\infty$.
For each $\beta_1$ given in Eq.~(\ref{beta1}) and $j\geq 2$,
the two conditions $\left<{\bp}_*,{\bf B}_j\right>= 0$ and $\left<{\bs_*},{\bf B}_{j+1}\right>= 0$
give rise to a real linear system for $k_j$ and $\beta_j$ which is uniquely solvable 
and guarantees that Eq.~(\ref{jj}) has a solution $\bp_j$ decaying at infinity and satisfying Eq.~(\ref{symort}).
Therefore, if $\chi(F)>0$ and $\delta>0$, we have two BICs in the perturbed waveguide. 

On the contrary, if $\chi(F)<0$, $\beta_1$ is complex, thus the perturbed waveguide (with $\delta>0$) does not have any BIC given as the power series~(\ref{Asthree}).
For perturbed waveguides with a negative $\delta$, the results can be obtained by substituting $\delta$ and $F$ with $-\delta$ and $-F$, respectively.

Notice that if $\delta$ is regarded as a parameter, two BICs emerge at $\delta=0$ [for $\delta>0$ or $\delta<0$, depending on the sign of $\chi(F)$].
Therefore, $\delta=0$ (corresponding to the non-generic BIC) is a bifurcation point.
Conversely, as $\delta$ tends to 0, these two BICs merge to the non-generic BIC.
This implies that the non-generic BIC is actually a {\em merging}-BIC when $\delta$ is the tuning parameter.
Existing studies on {\em merging}-BICs are concerned with specific examples and specific parameters \cite{koshiba08OL,Bulgakov17PRATopo,Zhen19Nature,Kang21PRL,kang22LSA,Bezus20Nano}.
Our study reveals that a non-generic BIC is a {\em merging}-BIC with respect to any general perturbation. 

\begin{figure}[h]
	\centering
	\includegraphics[width=6cm]{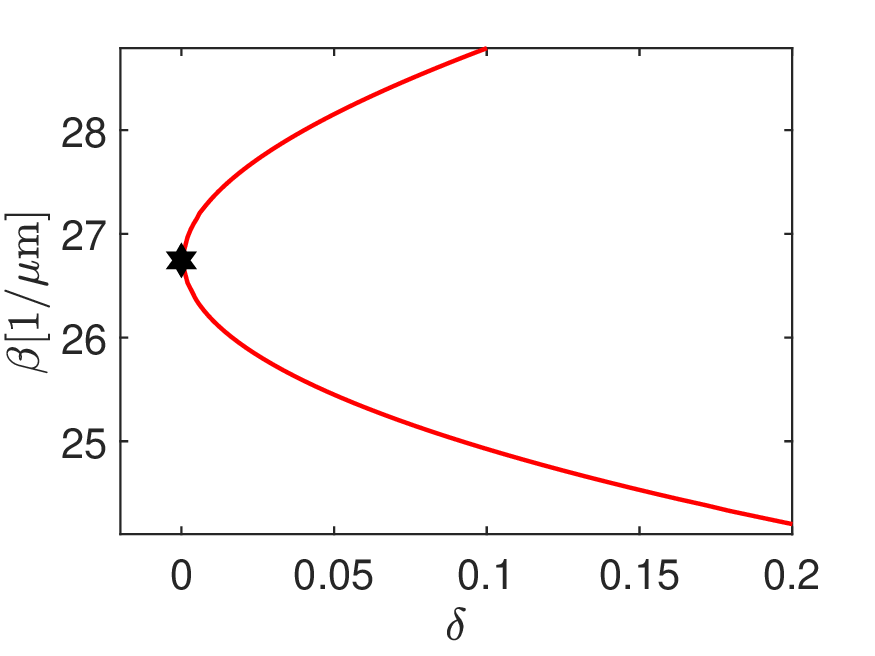}\quad
	\includegraphics[width=6cm]{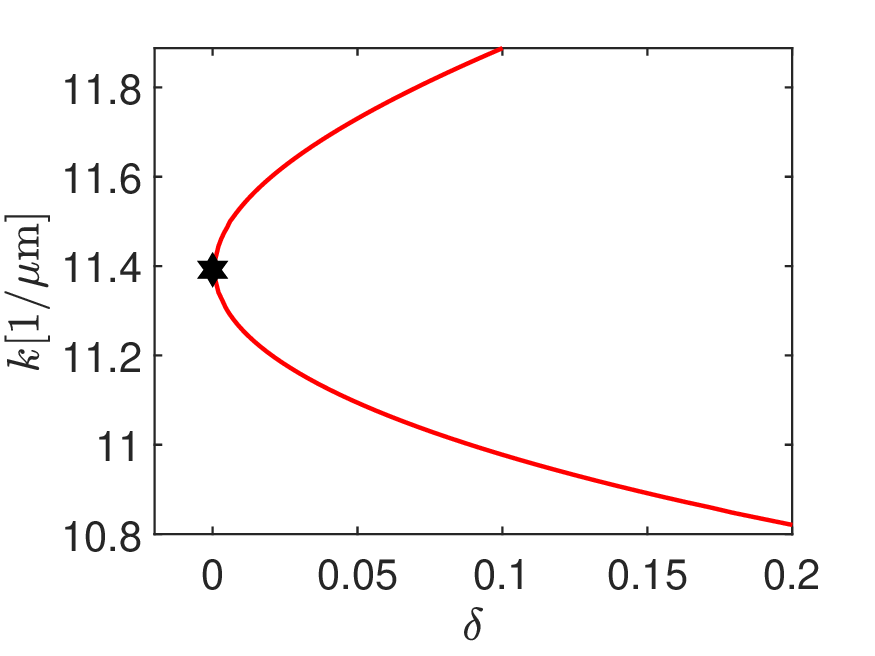}
	\caption{$\beta$ and $k$ of BICs emerging from a non-generic BIC marked by a black hexagram.}\label{Bifurcation}
\end{figure}

To verify our theory, we regard the silicon rib waveguide with $w=w_\natural$, studied in subsection~3.3, as the unperturbed waveguide.
In the following, we change the dielectric constant of the ridge and show the bifurcation phenomenon near the non-generic BIC.
More specially, we let the perturbation profile $F$ satisfy $F= -1$ and $F= 0$ in and outside the ridge, respectively.
For such a profile $F$, we can verify that $\chi(F)>0$. 
As shown in Fig.~\ref{Bifurcation}, for $\delta>0$,
two BICs emerge from the non-generic BIC at $\delta=0$ with the local behavior $\beta-\beta_*=\mathcal{O}(\sqrt{\delta})$ and
$k-k_*=\mathcal{O}(\sqrt{\delta})$.
For $\delta<0$, there is no BIC. 
On the other hand, if we assume that $F= 1$ and $F= 0$ in and outside the ridge respectively, we have $\chi(F)<0$. Therefore, two BICs exist in the perturbed waveguide with $\delta<0$ and no BIC exists for $\delta>0$. 
\section{Conclusion}

In this paper, we built a theoretical framework for non-generic BICs in waveguides with lateral leakage channels.
The definition of non-generic BICs is associated with the robustness theory developed in Ref. \cite{Yuan21OE}.
The generic and non-generic BICs are defined by a special integral which is non-zero and zero, respectively.
We developed a perturbation theory for resonant and leaky modes near generic and non-generic BICs.
It is shown that for a non-generic BIC with a real propagation constant $\beta_*$ and a frequency $\omega_*$, 
we typically have $Q\sim |\beta-\beta_*|^{-4}$ for the resonant mode with a real propagation constant $\beta$ near $\beta_*$. 
BICs surrounded by resonant modes with an ultra-high $Q$ factor have been
found in many works and they are referred to as  
{\em super}-BICs by some authors \cite{Yuri21NC,Bulgakov23PRB}.
Such a special BIC is usually obtained by merging a few BICs in a single dispersion surface/curve 
through tuning a structural parameter, and it is also referred to as a {\em merging}-BIC.
However, existing studies on {\em super}-BICs or {\em merging}-BICs are concerned with specific examples and specific parameters.
We studied general perturbations to waveguides supporting non-generic BICs, 
and developed a bifurcation theory for BICs in the perturbed waveguide.
Our work establishes interesting links among non-generic BICs, {\em super}-BICs and {\em merging}-BICs.
Notice that non-generic BICs are defined for the unperturbed waveguide, while {\em super}-BICs or {\em merging}-BICs
are related to perturbing or tuning of parameters.
Therefore, the existence of a non-generic BIC is an intrinsic property of the waveguide.

\section*{Appendix A}	
In this appendix, we show that $\mbox{Im}(k_2)$ defined in subsection 3.3 is proportional to $-|d_1|^2$.
Moreover, if $d_1=0$, we can prove that $\mbox{Im}(k_4)$ is proportional to $-|d_2|^2$.
In addition, the physical significance of $\mbox{Im}(k_2)$ is stated that it is associated with the leading-order radiation loss in the lateral direction.

Substituting Eqs. (\ref{Ask})-(\ref{AsE}) into Eq. (\ref{reducewaveeq}) and collecting $O(\delta^j)$ terms, we have
\[
\L\bp_j={\bf R}_j:=\B\bp_{j-1}/L+\2e\times\2e\times\bp_{j-2}/L^2+\sum_{l=1}^{j}\sum_{n=0}^{l}k_nk_{l-n}\varepsilon\bp_{j-l},\;j\geq 2,
\]
where $\bp_0=\bp_*$ and $k_0=k_*$.
To derive the imaginary part of $k_2$, we first recall some fundamental formulas.
We have the vector Green's theorem in 2D domain $\Omega$:
$$
\begin{aligned}
	\iint_\Omega\left(\mathbf{u} \cdot \nabla \times \nabla \times \mathbf{v}-\mathbf{v} \cdot \nabla  \times \nabla \times \mathbf{u}\right)\dr &=\int_{\partial \Omega}\left(\mathbf{v} \times \nabla \times \mathbf{u}-\mathbf{u} \times \nabla \times \mathbf{v}\right) \cdot {\bf n}\, \rmd \Gamma,\\
	\iint_\Omega\left(\mathbf{u} \cdot \nabla \times  \mathbf{v}-\mathbf{v} \cdot \nabla  \times\mathbf{u}\right)\dr &=\int_{\partial \Omega}({\bf n}\times {\bf v}) \cdot {\bf u} \,\rmd \Gamma,
\end{aligned}
$$
where $\bf n$ is the outer unit normal vector of $\partial\Omega$, ${\bf u}(\br)$ and ${\bf v}(\br)$ are vector functions. 
Recall the vector identities
$$
\begin{aligned}
	\nabla \cdot(\mathbf{a} \times \mathbf{b})&=(\nabla \times \mathbf{a}) \cdot\mathbf{b}-(\nabla \times \mathbf{b})\cdot\mathbf{a},\\
	\mathbf{a} \cdot(\mathbf{b} \times \mathbf{c})&=\mathbf{b} \cdot(\mathbf{c} \times \mathbf{a})=\mathbf{c} \cdot(\mathbf{a} \times \mathbf{b}).
\end{aligned}
$$
Because $\bp_*$ decays to zero exponentially as $|\br|\rightarrow 0$, by using the vector Green's theorem and vector identities, we have 
$
\left<\bp_*,\B\bp_1\right>=\left<\B\bp_*,\bp_1\right>.
$
The solvability of Eq.~(\ref{LphijEqu}) with $j=2$, i.e., $\left<\bp_*,{\bf R}_2\right>=0$, gives rise to
$$
2k_*\mbox{Im}(k_2)=-\mbox{Im}\left(\left<\B\bp_*,\bp_1\right>/L+2k_*k_1\left<\varepsilon{\bp}_*,\bp_1\right>\right)=-\mbox{Im}\left<{\L \bp_1},\bp_1\right>.
$$
According to the vector Green's theorem and vector identities,
we can obtain
$$
\begin{aligned}
	2i\mbox{Im}\left<{\L \bp_1},\bp_1\right>L^2=&\lim_{{\rm H}\rightarrow\infty}\iint_{\Omega_{\rm H}} \left(\overline{\L\bp}_1\cdot\bp_1-\cbp_1\cdot\L\bp_1\right)\,{\dr}\\
	=&\lim_{{\rm H}\rightarrow\infty}\int_{\partial \Omega_{\rm H}}\left[\cbp_1 \times \left(\nabla+i\beta_*\2e\right)\times\bp_1- \bp_1 \times \left(\nabla-i\beta_*\2e\right)\times \cbp_1\right]\cdot {\bf n}\rmd \Gamma\\
	=&4iL\alpha_{1*}^{\mathrm{te}}|d_1|^2,
\end{aligned}
$$
where $\Omega_{\rm H}=(-{\rm H},{\rm H})\times\mathbb{R}$ 
and $\alpha_{1*}$ is defined in Eq. (\ref{u}) with corresponding quantities.
Then we can get $\mbox{Im}(k_2)=-\alpha_{1*}^{\rm te}|d_1^2|/(k_*L)$.
Therefore, if the BIC is non-generic, we have $d_1=0$ and $\mbox{Im}(k_2)=0$. 
In this case, $\bp_1\rightarrow 0$ and ${\bf R}_2\rightarrow 0$ as $|\br|\rightarrow\infty$. It is clear that $\P{\bf R}_2={\bf R}_2$
and then Eq.~(\ref{LphijEqu}) with $j=2$ has a particular solution $\bp_2$ which satisfies $\P\bp_2 =\bp_2$ and has the following asymptotic form
\begin{equation}
	\bp_2 \sim d_2 {\bf u}_*^{\pm}, \quad x \to \pm \infty.
\end{equation}
The coefficient $d_2$ is a multiple of $\left<\bs_*,{\bf R}_2\right>$.
By using the same process as in the above, we can prove that $\mbox{Im}(k_4)$ is proportional to $-|d_2|^2$.

The complex Poynting vector ${\S}$ for the resonant modes near BICs can be expanded as
$$
{\S}=\frac{1}{2Z_0}{\bf E}\times\overline{\bf H}={\S}_*+\delta{\S}_1+\delta^2{\S}_2+\cdots,
$$
where $\bE=\bp(\br) e^{i\beta y}$, $\nabla \times {\bE}=ik{\bf H}$, and $Z_0$ is the freespace wave impedance. 
We can show that
\begin{equation}\label{Poynting1}
	\lim_{{\rm H}\rightarrow\infty}\int_{-\infty}^\infty\mbox{Re}({\S}_{1x}|_{x={\rm H}}-{\S}_{1x}|_{x=-{\rm H}})\rmd z=0,
\end{equation}
and
\begin{equation}\label{Poynting2}
	-\mbox{Im}(k_2)
	=\frac{Z_0}{L^2}\lim_{{\rm H}\rightarrow\infty}\int_{-\infty}^\infty\mbox{Re}({\S}_{2x}|_{x={\rm H}}-{\S}_{2x}|_{x=-{\rm H}})\rmd z.
\end{equation}
Equations (\ref{Poynting1})-(\ref{Poynting2}) imply that the imaginary part of $k_2$ is associated with the leading-order radiation loss in the lateral direction.

In subsection 3.2, we use the perturbation theory to analyze leaky modes near BICs. 
Substituting Eqs.~(\ref{Asg})-(\ref{AsE2}) into
Eq.~(\ref{reducewaveeq}) and collecting $O(\delta^j)$ terms, we obtain 
\[
\L\bp_j={\bf L}_j:=2k_*\varepsilon\bp_{j-1}/L+\varepsilon\bp_{j-2}/L^2+\sum_{l=1}^{j}\left(\beta_l\B+\sum_{n=1}^{l-1}\beta_n\beta_{l-n}\2e\times\2e\times\right)\bp_{j-l},\;j\geq 2,
\]
where $\bp_0=\bp_*$ and $\beta_0=\beta_*$.

\section*{Appendix B}
To describe our bifurcation theory clearly, we expand $k^2$ by
\begin{equation}\label{Aspk2app}
	k^2=\sum_{j=0}^\infty K_j\delta^{j/2},\;K_0=k_*^2,\;K_j=2k_*k_j+\sum_{l=1}^{j-1}k_lk_{k-l},\;j\geq 1.
\end{equation}	
Substituting Eqs.~(\ref{Asthree}) and (\ref{Aspk2app}) into Eq. (\ref{reducewaveeq}) and collecting terms of different powers of $\delta^{j/2}$, we can get
$$
\L\bp_j={\bf B}_j:=\beta_j\B\bp_*+K_j\varepsilon_*\bp_*+{\bf F}_j(\br),\;j\geq 2.
$$
The functions ${\bf F}_j$ are given by
\[
{\bf F}_2=\beta_1^2\2e\times\2e\times\bp_*+(\beta_1\B+K_1\varepsilon_*)\bp_1+k_*^2 F\bp_*,
\]
\[	
{\bf F}_j=\sum_{l=1}^{j-1}\left(\beta_l\B+K_l\varepsilon_*\right)\bp_{j-l}+\sum_{l=1}^{j}\left(K_{l-2}F+\sum_{n=1}^{l-1}\beta_n\beta_{l-n}\2e\times\2e\times\right)\bp_{j-l},\;j>2.
\]
Let $\hat{K}_1=-\left<\bp_*,\B\bp_*\right>$, then the relation between $\beta_1$ and $K_1$ can be written as
$K_1=\hat{K}_1\beta_1$.
Thus Eq. (\ref{j1}) becomes
$$
\L\bp_1=\beta_1\left(\B\bp_*+\hat{K}_1\varepsilon_*\bp_*\right).
$$
Since $\beta_1$ is unknown, $\bp_1$ cannot be solved, but it can be written as $\bp_1=\beta_1\hat{\bp}_1$, where
$\hat{\bp}_1$ satisfies
$\L\hat{\bp}_1=\B\bp_*+\hat{K}_1\varepsilon_*\bp_*.$
The function $\hat{\bp}_1$ can be scaled such that $\bp_1$ satisfies Eq. (\ref{symort}) if $\beta_1$ is real.
Using the above results, we can rewrite Eq. (\ref{jj}) with $j=2$ as
$$
\L\bp_2=\beta_2\B\bp_*+K_2\varepsilon_*\bp_*+\beta_1^2\hat{\bf R}_2+K_* F\bp_*,
$$
where 
$$\hat{\bf R}_2=\B\hat{\bp}_1+\2e\times\2e\times\bp_*+\hat{K}_1\varepsilon_*\hat{\bp}_1.$$
The condition $\left<{{\bf \Psi}_*},{\bf B}_2\right>=0$ gives rise to
$$
\beta_1^2A+k_*^2\left<F{{\bf \Psi}_*},\bp_*\right>=0,A=\left<{\bf \Psi}_*,\hat{\bf R}_2\right>.
$$
As shown in Appendix A, the coefficient $A$ is a multiple of $d_2$ since $\left<\varepsilon_*{\bf \Psi}_*,\bp_*\right>=0$.
In this paper, we assume that $d_2\neq 0$.
If $\chi(F)>0$, we have two real $\beta_1=\pm\sqrt{\chi(F)}$.
Accordingly, we can get a real $K_1$.
With $K_1$ and $\beta_1$ determined, Eq. (\ref{j1}) has a solution $\bp_1$ satisfying Eq. (\ref{symort}).

According to previous results, for each $\beta_1$ given in Eq. (\ref{beta1}) and $j\geq 2$, Eq. (\ref{jj}) can be written as
$$
\L\bp_j=\beta_j\B\bp_*+K_j\varepsilon_*\bp_*+2\beta_1\beta_{j-1}\hat{\bf R}_2+{\bf G}_j(\br),
$$
where ${\bf G}_j={\bf F}_j-2\beta_1\beta_{j-1}\hat{\bf R}_2$.
It is clear that ${\bf G}_j$ is independent of the unknowns $\beta_j$ and $K_j$.
The condition $\left<\bp_*,{\bf B}_{j}\right>=0$ gives rise to a real relation between $K_{j}$ and $\beta_{j}$:
\begin{equation}\label{apprelation}
	K_{j}-\hat{K}_1\beta_{j}=-\left<\bp_*,{\bf F}_{j}\right>.
\end{equation}
Although $\beta_{j}$ and $K_{j}$ are not obtained, we can reformulate $\bp_j$ as $\bp_j=\beta_j\hat{\bp}_1+{\bf w}_j$, where
$\L{\bf w}_j={\bf F}_j-\left<\bp_*,{\bf F}_{j}\right>\varepsilon_*\bp_*$.
Since the solution ${\bf w}_j$ is not unique, we can scale ${\bf w}_j$ such that $\bp_j$ satisfies Eq. (\ref{symort}) if $\beta_j$ is real.
Using above results, the condition $\left<\bs_*,{\bf B}_{j+1}\right>=0$ gives rise to
a real linear equation of $\beta_{j}$:
\begin{equation}\label{appbetaj}
	2A\beta_1\beta_{j}=-{\left<\bs_*,{\bf G}_{j+1}\right>}.
\end{equation}
Equations (\ref{apprelation}) and (\ref{appbetaj}) determine real $\beta_{j}$ and $K_{j}$ corresponding to each $\beta_1$ given in Eq.~(\ref{beta1}). 
With $K_j$ and $\beta_j$ determined, Eq.~(\ref{jj}) has a particular solution
$\bp_{j}$ satisfying Eq. (\ref{symort}).

\bibliography{ArXivNongeneric}

\end{document}